\documentclass[12pt]{article}
\usepackage{amstex}
\renewcommand{\section}[1] {\vspace{0.6cm}\addtocounter{section}{1}
\setcounter{subsection}{0}\setcounter{subsubsection}{0}
\noindent{\bf\thesection. #1}\par\vspace{0.4cm}}
\renewcommand{\subsection}[1] {\vspace{0.6cm}\addtocounter{subsection}{1}\setcounter{subsubsection}{0}\noindent{\it\thesubsection. #1}\par\vspace{0.4cm}}
\renewenvironment{thebibliography}[1]
{\vspace {0.6cm}\noindent {\normalsize\bf References}
\small\baselineskip=12.8pt\begin{list}
{[\arabic{enumi}]}
{\usecounter{enumi}\setlength{\parsep}{0pt}
\setlength{\leftmargin 1.25cm}{\rightmargin 0pt}
\setlength{\leftmargin 0.75cm}{\rightmargin 0pt}
 \setlength{\itemsep}{3pt} \settowidth
{\labelwidth}{#1.}\sloppy}}{\end{list}}

\begin{document}

\title{Finite-Difference Investigation of Axisymmetric Inviscid Separated Flows
with Infinitely-Long Cusp-Ended Stagnation Zone. Flow around a Sphere}
\author{M. D. Todorov \\
{\small \textit{Dept. of Differential Equations, Institute of Mathematics
and Informatics, }}\\
[-1.mm] {\small \textit{Technical University of Sofia, Sofia 1756, Bulgaria}}%
\\
[-1.mm] {\small \textit{e-mail: mtod@@vmei.acad.bg}}}
\date{ }
\maketitle

\begin{abstract}
The classical Helmholtz problem is applied for modelling the axisymmetric
inviscid cusp-ended separated flow around a sphere. Two coordinate systems
are employed: polar for initial calculations and parabolic the latter being
more suitable for investigation of infinitely long stagnation zones. Scaled
coordinates are introduced and difference schemes for the free-stream
equation and the Bernoulli integral are devised. The separation point is not
initially prescribed and is defined iteratively. A separated flow with
vanishing drag coefficient is obtained.
\end{abstract}

\section{Introduction}

In an attempt to explain the existence of a sizable drag force upon a
submerged body even for vanishing viscosity, Helmholtz \cite{helmholtz}
introduced the notion of discontinuous ideal flow consisting of a potential
and stagnant parts; these matching at an unknown stream surface. The idea of
discontinuous ideal flow was successfully applied by Kirchhoff \cite
{kirchhoff} for bodies with sharp edges and later developed by Levi-Civita
\cite{levicivi}, Villat \cite{villat}, Brodetsky \cite{brode}, etc. for
bodies with curved profile when additional condition for smooth separation
(Brillouin-Villat condition) is to be satisfied. All these solutions are
planar and based on the hodograph method. Unfortunately this powerful tool
is not capable for ideal flows characterized by axial symmetry. Therefore
the efforts in solving of such kind flows is mainly confined to the
numerical approach. Hitherto there are known several approximate methods for
study of axisymmetric ideal flows. The most important methods appear to be:
the integral one used at first by Trefftz \cite{trefftz} and later extended
by Struck \cite{struck}; the relaxation one applied by Southwell\&Vaisey
\cite{soutvaisy} and developed by Brennen \cite{brennen69} (for detailed
reference see \cite{wu72, gurevich}). Similarly to the plane flows in the
case of curve bodies a smooth separation condition or any else
semi-empirical assumptions are suggested in order to yield satisfactory
forecast concerning the velocity and pressure distribution, detachment point
and drag coefficient \cite{armstr, armstrdunh,plesshaf}. Particularly
Southwell\&Vaisey by working in the physical plane obtained only cusp-ended
cavity behind a sphere. We also calculated such kind stagnation zone behind
a sphere \cite{todo91} by means of finite-difference scheme at that without
pre-conditioning the separation point. Now we aim at utilizing the improved
difference scheme, which was developed and applied for the planar inviscid
flow around circular cylinder \cite{tod98} for investigation of a separated
axisymmetric flow around a sphere.

Following our approach we use two different coordinate systems: a polar
spherical
coordinate system for initial calculations and a parabolic coordinate system
the latter being topologically more suited for solving the free-stream
equation outside infinitely-long stagnation zones. We switch from polar
coordinates to parabolic ones after the stagnation zone has fairly well
developed and has become long enough.

\setcounter{equation}{0}

\section{Posing the Problem}

Consider the steady inviscid flow past a circle -- an arbitrary meridian
cross section of a sphere. The direction of the flow coincides with the line
$\theta =0,\pi $ of the polar coordinates and the leading stagnation point
of the flow is situated in the point $\theta =\pi $. The axially symmetry
enables to study the flow in the meridian halfplane only.

Dimensionless variables are introduced as follows
\begin{gather}
\psi^{\prime}= {\frac{\psi }{{L^2 U_\infty}}}, \ \ r^{\prime}= {\frac{r }{L}}%
, \ \ q = {\frac{{p-p_c} }{{{\frac{1 }{2}} \rho U^2_\infty}}}, \ \ \sigma =
\sqrt{L}\sigma^{\prime}, \ \ \tau = \sqrt{L} \tau^{\prime}, \ \ \kappa = {%
\frac{{p_\infty - p_c} }{{{\frac{1 }{2}} \rho U^2_\infty}}},
\label{eq:nondim}
\end{gather}

\noindent where $L$ is the characteristic length of the body ($2a$ for a
sphere of radius $a$), $U_{\infty }$ -- velocity of the undisturbed flow; $%
p_{c}$ -- the pressure inside the stagnation zone; $p_{\infty }$ -- the
pressure at infinity, $r$ - the polar radius, $\sigma ,\tau $-the parabolic
coordinates, $\kappa$ - the cavitation number,
which for flows with stagnation zones is equal to zero. Without fear of
confusion the primes will be omitted henceforth.

\subsection{Coordinate Systems}

In terms of the two coordinate systems (polar spherical and parabolic) equation
for the stream function $\psi$ reads:
\begin{equation}
{\frac{1 }{\sin \theta}} (\psi_r)_r + {\frac{1 }{r^2}} \left({\frac{%
\psi_\theta }{\sin \theta}}\right)_\theta = 0 \>, \qquad\hbox{or}\qquad
\frac{1}{\tau} \left( \psi_\sigma + \frac{\psi}{ \sigma} \right)_\sigma +
\frac{1}{\sigma} \left( \psi_\tau + \frac{\psi}{ \tau} \right)_\tau = 0\>.
\label{eq:govern}
\end{equation}

The undisturbed uniform flow at infinity is given by
\begin{equation}
\left. \psi \right|_{r \to \infty} \approx {\frac{r^2 U_\infty \sin^2{\theta}
}{2}} \>, \qquad \hbox{or} \qquad \left. \psi\right|_{\sigma \to \infty, \>
\tau \to \infty} \approx {\sigma \tau U_\infty } \>.  \label{eq:integbc}
\end{equation}

On the combined surface ``body+stagnation zone'' hold two conditions. The
first condition secures that the said boundary is a streamline (say of
number ``zero'')
\begin{equation}
\psi(R(\theta),\theta) =0, \> \theta \in [0,\pi] \quad \hbox{or} \quad
\psi(S(\tau),\tau) = 0, \> \tau \in (0,\infty) \>,  \label{eq:psizero}
\end{equation}

\noindent where $R(\theta )$, $S(\tau )$ are the shape functions of the
total boundary in spherical or parabolic coordinates, respectively. As
usually
we use the notation $\Gamma _{1}$ for the portion of boundary representing
the rigid body (the sphere) and $\Gamma _{2}$ -- for the free streamline
(Fig.1).

On $\Gamma _{2}$ the shape function $R(\theta )$ is unknown
and it is to
be implicitly identified from Bernoulli integral with the pressure equal to a
constant (say, $p_{c}$) which is the second condition holding on the free
boundary. For the two coordinate systems one gets the following equations
for shape functions $R(\theta )$ or $S(\tau )$:
\begin{eqnarray}
\left[ q+\frac{1}{r^{2}\sin ^{2}\theta }(\frac{\psi _{\theta }^{2}}{r^{2}}%
+\psi _{r}^{2})\right] _{r=R(\theta )}=1\>, &\qquad \hbox{or}\qquad &\left[
q+\frac{\psi _{\sigma }^{2}+\psi _{\tau }^{2}}{\sigma ^{^{2}}+\tau ^{^{2}}}%
\right] _{\sigma =S(\tau )}=1\>.  \label{eq:dyncond} \\
\theta \in \Gamma _{2}\qquad\qquad\qquad\qquad\qquad &&\quad\quad \tau \in \Gamma _{2} \nonumber
\end{eqnarray}

The boundary value problem (\ref{eq:govern}), (\ref{eq:integbc}), (\ref
{eq:psizero}), (\ref{eq:dyncond}) is completed with the  additional symmetry
conditions
\begin{equation}
\frac{\partial \psi}{\partial \theta} = 0 \>, \ \theta = 0, \pi \qquad%
\hbox{or}\qquad \frac{\partial \psi}{\partial \tau} = 0 \>, \ \tau = 0 \>.
\label{eq:symconds}
\end{equation}

In spherical coordinates along with $\psi $ it	is convenient to introduce new
function $\Psi =\frac{\psi }{r\sin \theta }$. Then the dynamical condition (%
\ref{eq:dyncond}a) takes the form:
\begin{eqnarray}
\left[ q+\frac{\Psi _{\theta }^{2}}{r^{2}}+\Psi _{r}^{2}\right] _{r=R(\theta
)} &=&1\>.  \label{eq:dyncond1} \\
\theta \in \Gamma_2 \qquad\qquad&& \nonumber
\end{eqnarray}
Obviously $\left. \Psi _{r}\right| _{r=R(\theta )}=\left. \frac{\psi _{r}}{%
r\sin \theta }\right| _{r=R(\theta )}\>,\>\left. \Psi _{\theta }\right|
_{r=R(\theta )}=\left. \frac{\psi _{\theta }}{r\sin \theta }\right|
_{r=R(\theta )}$. Without confusion we will name $\Psi $ stream function too.

\subsection{Scaled Variables}

Following \cite{chritodo84,chritodo86,chritodo87} we introduce new scaled
coordinates:
\[
\eta =rR^{-1}(\theta )\>,\qquad \eta =\sigma -S(\tau ),
\]
which render the original regions to semi-infinite strips.

If we denote $\xi \equiv \theta $ or $\xi \equiv \tau $ depending on the
particular case under consideration, then in terms of the new coordinates $%
(\eta ,\xi )$, the governing equation (\ref{eq:govern}) takes the form
\begin{equation}
A(\psi _{\eta })_{\eta }+B(b\psi _{\xi })_{\xi }-C(\psi _{\xi })_{\eta
}-D(d\psi _{\eta })_{\xi }+(e\psi )_{\eta }+(f\psi )_{\xi }=0\>,
\label{eq:laplace}
\end{equation}
\noindent where
\begin{eqnarray*}
b \equiv \frac{1}{\sin \theta} \>, \quad d \equiv {\frac{R^{\prime}}{R}} {%
\frac{1 }{\sin \theta}} \>,&& \!\!\!\! e \equiv 0\>, \quad f \equiv 0; \\
A \equiv \eta^2 + \eta {\left (\frac{R^\prime}{R} \right)}^2\>,\quad B
\equiv \sin \theta \>, && \!\!\!\! C \equiv \eta \frac {R^{\prime}}{R} \>, \quad D
\equiv \eta \sin \theta \>;
\end{eqnarray*}
\begin{center} or \end{center}
\begin{eqnarray*}
b \equiv 1 \>, \quad d \equiv S^{\prime}\>, && \!\!\!\! e \equiv \frac{1}{\eta + S}
- \frac{S^\prime}{\tau}\>, \quad f \equiv \frac{1}{\tau}; \\
A \equiv 1 + {S^{\prime}}^2 \>, && \!\!\!\! B \equiv 1 \>, \quad C \equiv
S^{\prime}\>, \quad D \equiv 1\>.
\end{eqnarray*}

Similarly to \cite{tod98} we use the ``relative'' function $\bar\psi$
\[
\bar\psi(\eta, \theta) = \psi (\eta, \theta) - {\frac{[\eta R(\theta) \sin
\theta]^2 }{2}} \>, \quad \bar\psi(\eta, \tau) = \psi (\eta,\tau)- {(\eta
+S(\tau)) \tau } \>,
\]

\noindent which is obviously a solution to eq.(\ref{eq:laplace}) and which
we loosely call stream function. The asymptotic boundary condition then
becomes
\begin{equation}
\left. \bar\psi \right|_{\eta= \eta_\infty} = 0 \qquad \hbox{or} \qquad
\left. \bar\psi \right|_{\eta=\eta_\infty,\> \tau = \tau_\infty} = 0 \>,
\label{eq:inftybc}
\end{equation}

\noindent while the non-flux condition on $\Gamma$ transforms as follows
\begin{equation}
\bar\psi \big|_{\eta=1} = - {\frac{[R(\theta) \sin \theta]^2 }{2}} \qquad %
\hbox{or} \qquad \bar\psi \big|_{\eta=0} = - {S(\tau) \tau } \>.
\label{eq:bodybc}
\end{equation}

Thus eqs.(\ref{eq:laplace}), (\ref{eq:inftybc}), (\ref{eq:bodybc}), (\ref
{eq:symconds}) define a well posed boundary value problem provided that
functions $R(\theta )$ and $S(\tau )$ are known. On the other hand in the
portion $\Gamma _{2}$ of the boundary (where these functions are unknown)
they can be evaluated from the Bernoulli integral (\ref{eq:dyncond}) and (%
\ref{eq:dyncond1}) which now becomes an explicit equation for the shape
function
\begin{eqnarray}
{\frac{R^{2}+{R^{\prime }}^{2}}{R^{6}\sin ^{2}\theta }}\left[ \left. {\frac{%
\partial \bar{\psi}}{\partial \eta }}\right| _{\eta =1}\!\!+R^{2}(\theta
)\sin ^{2}\theta \right] ^{2} &=&1
\nonumber \\ [-0.2cm] &&\qquad 0\le \theta \le \theta^* \nonumber\\[-0.2cm]
{\frac{R^{2}+{R^{\prime }}^{2}}{R^{4}}}\left[ \left. {\frac{\partial \bar{%
\Psi}}{\partial \eta }}\right| _{\eta =1}\!\!+R(\theta )\sin \theta \right]
^{2} &=&1\>,	    \nonumber \\[-0.2cm]
&&\hbox{or}  \label{eq:bernoulli} \\[-0.2cm]
{\frac{1+{S^{\prime }}^{2}}{S^{2}+\tau ^{2}}}\left[ \left. {\frac{\partial
\bar{\psi}}{\partial \eta }}\right| _{\eta =0}\!\!+\tau \right] ^{2}
&=&1\>,\quad \tau^* \le \tau <\infty \>.  \nonumber
\end{eqnarray}

Here $\bar\Psi(\eta, \theta) = \Psi (\eta, \theta) - \eta R(\theta) \sin
\theta$.

\setcounter{equation}{0}

\section{Forces Exerted on the Body}

Apparently the presence of a stagnation zone breaks the symmetry of the
integral for the normal stresses and hence D'Alembert paradox is not hold.
If denote by $\mbox{\boldmath $n$}$ the outward normal vector to the sphere $%
\Sigma $ and by $d\sigma $ - the surface element of the sphere, then the
force acting upon the body is given by
\begin{equation}
\mbox{\boldmath $R$}=-\oint_{\Sigma }p\mbox{\boldmath $n$}d\sigma
\label{eq:force_integral}
\end{equation}

It is not difficult to obtain for the drag and lifting-force coefficients of
every meridian cross section the following expressions
\begin{eqnarray}
C_{x}=-\int_{\theta^*}^{\pi }\!\!\!qR(\theta )\sin (\theta )\left[
R(\theta )\cos {\theta }+R^{\prime }(\theta )\sin {\theta }\right] d\theta  &%
\hbox{or}&C_{x}=\int_{0}^{\tau^*}\!\!\!qS(\tau )\tau \left[ S(\tau
)+S^{\prime }(\tau )\tau \right] d\tau	 \nonumber  \label{eq:cx_cy} \\%
[-0.2cm]
&& \\[-0.2cm]
C_{y} &\equiv &0\>,  \nonumber
\end{eqnarray}

\noindent where the dimensionless pressure is given by
\begin{eqnarray}
q &=& 1 - {\frac{R^2+{R^{\prime}}^2}{R^4}} \left[\left. {\frac{\partial
\bar\Psi }{\partial \eta}} \right|_{\eta=1} \!\! + R \sin \theta \right]^2
\!\!  \nonumber \\[-0.2cm]
&\hbox{or}  \label{eq:pressure} \\[-0.2cm]
q &=& 1 - {\frac{1 + {S^{\prime}}^2}{S^2 + \tau^2}} \left[ \left. {\frac{%
\partial \bar\psi}{\partial \eta}}\right|_{\eta=0} \!\!+ \tau \right]^2
\!\!\!\!\>.   \nonumber
\end{eqnarray}

\setcounter{equation}{0}

\section{Difference Scheme and Algorithm}

\noindent\vspace{-0.6cm}\noindent

\subsection{Splitting scheme for the free-stream equation}

\nobreak\noindent The computational domain being infinite is reduced to
finite
one after appropriately choosing the ``actual infinities''. In order to take
into consideration the topological and dynamic features of the flow we
employ non-uniform mesh, which was presented in detail at \cite{tod98}.

Let us denote the spacings of the mesh by $h_{i+1}\equiv \eta _{i+1}-\eta
_{i}\>,\ i=1,\ldots ,M$ and $g_{j+1}\equiv \xi _{j+1}-\xi _{j}\>,\
j=1,\ldots ,N$. We solve the boundary value problem iteratively using
the method of splitting of operator. Upon introducing fictitious time we
render the equation to parabolic type and then employ the so-called scheme
of stabilising correction \cite{yanenko}. On the first half-time step we
have the following differential equations ($\Delta t$ is the time increment)
\begin{gather}
\frac{\psi _{ij}^{n+\frac{1}{2}}-\psi _{ij}^{n}}{\frac{1}{2}\Delta t}=B_{ij}{%
\Lambda _{2}(b\Lambda _{2}\psi ^{n+\frac{1}{2}})}_{ij}+A_{ij}{\Lambda
_{1}(\Lambda _{1}\psi ^{n})}_{ij}-C_{ij}{\Lambda _{1}(\Lambda _{2}\psi ^{n})}%
_{ij}  \nonumber \\
-D_{ij}{\Lambda _{2}(d\Lambda _{1}\psi ^{n})}_{ij}+\Lambda _{1}(e\psi
^{n})_{ij}+\Lambda _{2}(f\psi ^{n})_{ij}  \label{eq:frst_step}
\end{gather}
for $i=2,\cdots ,M,\ j=2,\ldots ,N$

The second half-time step consists in solving the following differential
equations
\begin{gather}
\frac{\psi^{n+1}_{ij}-\psi^{n+\frac{1}{2}}_{ij}}{\frac{1}{2}\Delta t}%
=A_{ij}\left({\Lambda_1(\Lambda_1 \psi^{n+1})}_{ij} - {\Lambda_1(\Lambda_1%
\psi^n)}_{ij}\right)  \label{eq:sec_step}
\end{gather}

\noindent for $i=2,\ldots,M, \ j=2,\ldots,N$. The last two equations (\ref
{eq:frst_step})-(\ref{eq:sec_step}) are completed with respective boundary
conditions \cite{chritodo86}. Here
\[
\Lambda_1 {(.)}_{ij} \equiv \frac{\partial }{\partial \eta}(.)_{ij} + O(h_i
h_{i+1})\>,
\]
\[
\Lambda_2 (.)_{ij} \equiv \frac{\partial }{\partial \xi}(.)_{ij} + O(g_j
g_{j+1})
\]
are the usual difference operators based on three-point patterns with
second order of approximation.

Thus the b.v.p. for the stream function is reduced to consequative systems
with sparse (three-diagonal) matrices, which are solved iteratively \cite
{chritodo86}.

Since the condition for numerical stability of the elimination is not
satisfied for all points of domain here a ``non-monotonous progonka'' (see
\cite{samnik,chreding}) is employed like at \cite{tod98}.

\subsection{Difference Approximation for the Free Boundary}

Following \cite{tod98} in the present work we use the dynamic condition (\ref
{eq:dyncond}) in spherical coordinates only, so that we present here just
the
relevant scheme in spherical coordinates. The equations
(\ref{eq:bernoulli})
can be resolved for the derivative $R^{\prime}(\theta)$ when the following
conditions are satisfied:
\begin{eqnarray}
&&Q(\theta) \stackrel{\mathrm{def}}{=} {\frac{R^4(\theta ) \sin^2 \theta }{%
T^2 }} > 1 \>, \ T = \left. \frac{\partial \bar\psi}{\partial \eta}
\right|_{\eta=1} \!\! + \left(R(\theta) \sin\theta\right)^2  \nonumber \\%
[-0.2cm]
\hbox{or}&&  \label{eq:posit_cond} \\[-0.2cm]
&&Q(\theta) \stackrel{\mathrm{def}}{=} {\frac{R^2(\theta ) }{\mathcal{T}^2 }}
> 1 \>, \ \mathcal{T} = \left. \frac{\partial \bar\Psi}{\partial \eta}
\right|_{\eta=1} \!\! + R(\theta) \sin\theta\>.  \nonumber
\end{eqnarray}

\noindent The above inequalities are trivially satisfied in the vicinity of
the rear-end stagnation point inasmuch as that for $\theta\rightarrow0$ one
has $T \to 0$ or $\mathcal{T} \to 0$ and hence ${\frac{R^4 \sin^2 \theta }{%
T^2}}\to\infty$ or ${\frac{R^2}{\mathcal{T}^2}}\to\infty$. The first
inequality, however, is indeterminated at the point $\theta = 0$ due to the
ratio $\frac{\sin \theta }{T(\theta)}$.

For the shape function $\hat{R}_{j}$ of free line is solved the following
difference scheme
\begin{eqnarray}
&&\hat{R}_{j-1}\!\!-\!\hat{R}_{j}=g_{j}{\frac{\hat{R}_{j}+\hat{R}_{j-1}}{2}}%
\sqrt{\frac{1}{2}\left[ \left( {\frac{({R_{j}^{\alpha }})^{2}\sin \theta _{j}%
}{T_{j}^{\alpha }}}\right) ^{2}+\left( {\frac{(R_{j-1}^{\alpha })^{2}\sin
\theta _{j-1}}{T_{j-1}^{\alpha }}}\right) ^{2}\right] -1}  \nonumber \\
\hbox{or} &&  \label{eq:fs_polar} \\[-0.2cm]
&&\hat{R}_{j-1}\!\!-\!\hat{R}_{j}=g_{j}{\frac{\hat{R}_{j}+\hat{R}_{j-1}}{2}}%
\sqrt{\frac{1}{2}\left[ \left( {\frac{R_{j}^{\alpha }}{\mathcal{T}%
_{j}^{\alpha }}}\right) ^{2}+\left( {\frac{R_{j-1}^{\alpha }}{\mathcal{T}%
_{j-1}^{\alpha }}}\right) ^{2}\right] -1}  \nonumber
\end{eqnarray}

\noindent for $j=j^*, \ldots ,2$ , whose approximation is $O(g^2_j)$. Only
in the detachment point the difference scheme is different, specifying in
fact the initial condition, namely
\begin{eqnarray*}
&&\hat R_{j^*} - R(\theta^*) = g^* {\frac{R(\theta^*)+ \hat R_{j^*}}{2}}
\sqrt{\frac{1}{2}\left[\left(\frac{(R^\alpha_{j^*})^2 \sin \theta_{j^*} }{%
T^\alpha_{j^*} }\right)^2 +\left({\frac{R^2(\theta^{*}) \sin \theta^*}{%
T(\theta^*)}}\right)^2\right] - 1} \\[-0.1cm]
\hbox{or}&& \\[-0.1cm]
&&\hat R_{j^*} - R(\theta^*) = g^* {\frac{R(\theta^*)+ \hat R_{j^*}}{2}}
\sqrt{\frac{1}{2}\left[\left({\frac{R^\alpha_{j^*} }{\mathcal{T}^\alpha_{j^*}%
}}\right)^2 +\left({\frac{R(\theta^{*})}{\mathcal{T}(\theta^*)}}%
\right)^2\right] - 1}\>,
\end{eqnarray*}

\noindent where $R$ without a superscript or ``hat'' stands for the known
boundary of rigid body.

At last a relaxation is used for the shape-function of the free boundary at
each global iteration $\alpha $ according to the formula:
\[
R^{\alpha +1}=\omega \hat{R}_{j}+(1-\omega )R_{j}^{\alpha }\>,
\]

\noindent where $\omega$ is called relaxation parameter.

\subsection{The general Consequence of the Algorithm}

\nobreak\noindent Each global iteration contains two stages. On the first
stage, the difference problem for free-stream equation is solved iteratively
either in polar spherical or in parabolic coordinates (depending on the
development of the stagnation zone).

The second stage of a global iteration consists in solving the difference
problem for the free surface in polar spherical coordinates.

Through the indetermination at the axis of symmetry we use the difference
scheme (\ref{eq:fs_polar}a) only during the first several iterations (in
polar spherical coordinates). The calculation of the shape of the far weak (in
parabolic coordinates) we carry out using the scheme (\ref{eq:fs_polar}b).
The latter appears to be more convenient and efficient because the loss of
accuracy and 'numerical' instability in vicinity of the axis of cusp are
avoided. The criterion for convergence of the global iterations is defined
by the convergence of the shape function, namely
\begin{equation}
\max_{j}\big| {\frac{R_{j}^{\alpha +1}-R_{j}^{\alpha }}{R_{j}^{\alpha +1}}}%
\big|<10^{-4}.	\label{eq:normR}
\end{equation}

The obtained solutions for the stream function and the shape function of the
boundary are the values of the last iteration $\psi_{ij} =
\psi^{\alpha+1}_{ij}$ and $R_j = R^{\alpha+1}_{j}$, respectively. Then the
velocity, pressure, and the forces exerted from the flow upon the body are
calculated.

\setcounter{equation}{0}

\section{Results and Discussion}

The numerical correctness of scheme (\ref{eq:frst_step}),
(\ref{eq:sec_step}) is verified
through usual experiments including  a doubling the mesh knots and varying
the 'actual infinity' We used different meshes with sizes $M$ x $N$ : 41x68,
81x136,  101x202, etc. Respectively, the actual infinity $\eta _{\infty }$
assumed in the numerical experiments the values 10, 20. The dependence of
the numerical solution on the time increment $\Delta t$ is also investigated
and it is shown that the scheme of fractional steps for the stream function
has a  full approximation \cite{yanenko}.
Comparing the
different finite-difference realizations of the solution we
choice the following 'optimal' values of the governing parameters: step
of the fictitious time $\Delta
t=0.5$, relaxation $\omega =0.01$ and 'actual' infinity $\eta _{\infty }=10$.

In Fig.2-a are presented the obtained shapes of the
stagnation zone behind the sphere and in the near wake for resolutions $%
41\times 68$, $81\times 136$ and $101\times 202$ and value of relaxation
parameter: $\omega = 0.01$.

Evidently the agreement among the calculated shapes of the free boundary
near the body corresponding to these three meshes is very well. The logarithmic
scale is used in Fig.2-b in order to expand the differences
between the different solutions making them visible in the graph.
As clearly it is shown the curves are indistinguishable till distance 150
calibers and the relative error is less than 1\%.  The relative
error between the meshes $81 \times 136$ and $101 \times 202$ at distances more
than 150 calibers does not exceed
4\%. At the same time the relative error between the mesh $41 \times 68$
and the else
two ones increases and reaches 7-8\% at distance 200 calibers. Obviously that
mesh is not enough fine and appears to be coarse for calculating the shape
function at large distances behind the sphere. The obtained
results warrant
conclusion that the scheme is fully effective in solving the free boundary
till 200 calibers.
The very good comparison supports the claim that indeed a solution to
the Helmholtz problem has been found numerically by means of the developed
in the present work difference scheme.
The calculated here dimensionless pressure $q$ is shown in Fig.3.
The agreement among the obtained pressure curves corresponding
to different mesh resolutions is excellent.
In the stagnation zone the pressure is in order of $10^{-4}$ in accordance with
the assumption that the unknown boundary is defined by the
condition $q=0$. The amplitude of the minimum of $q$ is smaller than $1.25$ the
latter being the value for ideal flow without separation. This means that
the stagnation zone influences the flow upstream. The calculated
magnitude of the separation
angle (measured with respect the rear end of the sphere) varies between
$69.42^\circ$  for mesh $41 \times 68$ and $69.7^\circ$ for mesh $101 \times
202$. It is interesting to note that the calculated here drag
coefficient $C_x$ has a magnitude between $.5848 \times 10^{-3} - .5704 \times
10^{-2}$
obtained for the different resolutions, i.e., we conclude that in order of
approximation of the scheme $C_x = 0$. Then similarly to the separated flow
around
circular cylinder we can name the obtained separation angle 'critical' (see
\cite{tod98,gurevich}). Hence in the case of axisymmetric flow around
sphere
there also exists a inviscid separated flow for which the D'Alembert paradox
holds.
Trough the disscused features of the obtained Helmholtz flow we can assume it
is an axisymmetric analogue of the Chaplygin-Kolscher flow around
circular cylinder.

\section{Concluding Remarks}

The separated inviscid flow behind a sphere is treated as a
flow with free surface -- the boundary of the stagnation zone (Helmholtz
problem). Scaled coordinates are employed rendering the computational domain
into a region with fixed boundaries and transforming the Bernoulli integral
into an explicit equation for the shape function. A new free-stream
function is introduced and thus the numerical instability near the
symmetry axis is avoided. Difference scheme using
coordinate splitting is devised. Exhaustive set of numerical experiments is
run and the optimal values of scheme parameters are defined. Results are
verified on grids with different resolutions. The obtained here shape of the
stagnation zone is of infinitely long cusp and respective separated flow has
vanishing drag coefficient.
The detachment point is not prescribed in
advance and it is defined iteratively satisfying the mere Bernoulli integral
there.

\bigskip \noindent \textbf{Acknowledgment\/} The author presents his gratitudes
to Prof. C.I.Christov for stimulation to carry out this research and useful
advices.

This work was
supported by the National Science Foundation of Bulgaria, under Grant
MM-602/96.

\bigskip \bigskip \centerline{\bf FIGURE CAPTIONS}

\bigskip \centerline{\it fig1.gif}

\centerline{Figure 1: Posing the problem}

\bigskip \centerline{\it sphnear.gif}

\centerline{(a) behind the sphere}

\bigskip \centerline{\it sphfar.gif}

\centerline{(b) far wake}

\noindent Figure 2: The obtained separation lines for relaxation parameter $%
\omega = 0.01$ and different resolutions: - - - - $41\times 68$; --- --- ---
$81\times 136$; -- -- -- $101\times 202$.

\bigskip \centerline{\it sphpres.gif}

\noindent Figure 3: The pressure distribution for relaxation parameter $%
\omega = 0.01$ and different resolutions: - - - - $41\times 68$; --- --- ---
$81\times 136$; -- -- -- $101\times 202$; ------ inviscid nonseparated flow.

\end{document}